\documentclass[aps,pra,amsmath,amssymb,twocolumn]{revtex4}

\usepackage[T1]{fontenc}
\usepackage{amssymb,amsmath}
\usepackage{amsfonts}
\usepackage{graphicx}
\usepackage{amsthm}
\usepackage{color}
\usepackage{enumerate}
\usepackage{mathbbol}

\def\vec#1{\mathbf{#1}}
\def\ket#1{|#1\rangle}

\def\bra#1{\langle#1|}

\def\tr{\mathrm{tr}}

\def\bbc{\mathbb{C}}

\newtheorem{theorem}{Theorem}[]

\usepackage{listings}

\newcommand{\bfmu}{\boldsymbol{\mu}}
\newcommand{\bfnu}{\boldsymbol{\nu}}

\renewcommand{\leq}{\leqslant}
\renewcommand{\geq}{\geqslant}

\begin{document}

\title{Entanglement and the truncated moment problem}

\author{F.~Bohnet-Waldraff$^{1,2}$, D.~Braun$^{1}$, and O.~Giraud$^{2}$}
\affiliation{$^{1}$Institut f\"ur theoretische Physik, Universit\"at T\"{u}bingen,
72076 T\"ubingen, Germany\\
$^{2}$\mbox{LPTMS, CNRS, Univ.~Paris-Sud, Universit\'e
    Paris-Saclay, 91405 Orsay, France}
}

\begin{abstract}
We map the quantum entanglement problem onto the mathematically
well-studied truncated moment problem. This yields a necessary and
sufficient condition for separability that can be checked
by a hierarchy of semi-definite programs. The algorithm always gives a
certificate of entanglement if the state is entangled. If the state is
separable, typically a certificate of separability is obtained in a
finite number of steps and an explicit decomposition into separable
pure states can be extracted.    
\end{abstract}
\date{April 7, 2017}
\maketitle

\section{Introduction}
\label{sec:intro}

The renewed interest that entanglement theory attracted in the last
decades has led to a tremendous amount of new results (see the recent
reviews  \cite{Bruss01,Plenio05,Hor09,Guehne2009} and
references therein). 
Still, characterization and detection of multipartite entanglement is largely an open question. For quantum states describing a collection of qubits, the size of Hilbert space, exponential in the number of qubits, makes the problem daunting.  A simpler but still challenging problem is to restrict the question of characterizing entanglement to a smaller set of quantum states, such as for instance symmetric states, which are pure states invariant under permutations of constituents, or mixtures thereof. Symmetric states lie in a Hilbert space of size linear in the number of qubits, which makes the investigation more tractable. Once the symmetric case is understood, it can shed light onto the general case. This is the strategy we will follow here, first considering the symmetric case, which is easier to handle and to present from a pedagogical point of view, then extending our results to the fully general non-symmetric case.

Various results on entanglement for symmetric states have been
obtained in the literature \cite{Dev07, Hub09, Tot09, Aug12,
  Bag14}. For instance, criteria for certifying separability in symmetric
mixed states of $N$ qubits were found in \cite{WolYel14}. Separable
symmetric $N$-qubit pure states are always fully separable
\cite{Itchikawa}. 
They are 
easily characterized, as there is a one-to-one correspondence between
these states and points on the Bloch sphere via the Majorana
representation \cite{Maj32}. As will be detailed in the paper, a
symmetric state is separable (that is, it can be written as a convex
combination of separable pure states) if and only if it can be
associated with a probability distribution on the sphere; this measure
then gives the positive weights associated with each separable pure
state. A convenient representation to describe symmetric states in
terms of symmetric tensors was proposed in \cite{PRL_Tensor},
generalizing the Bloch sphere picture of spins-1/2. In terms of this
tensor representation, decomposing a state into a convex combination
of separable pure states amounts to finding a probability distribution
whose lowest-order moments are fixed by the tensor entries. In fact,
as we will see, the generic problem of finding whether an arbitrary
(not necessarily symmetric) multipartite state can be decomposed into
product states can be cast into the 
problem of finding a probability distribution whose lowest-order
moments are fixed.

The problem of finding a probability distribution from the knowledge of its moments has been extensively studied in the literature. When only a finite number of moments is known, the problem is to find {a} probability distribution compatible with these moments. In the case of multivariate distributions, it corresponds to the so-called {\it truncated moment problem}: given a truncated moment sequence (tms), that is, fixing all moments up to a certain order, is there a probability distribution (or, in mathematical terms, a nonnegative measure) whose moments coincide with those of the tms? When it exists, such a measure is called a representing measure of the tms. Of practical relevance is the closely related $K$-tms problem, where the measure reproducing the fixed moments is constrained to be supported on some compact $K$. 

The non-truncated $K$-moment problem, where all moments are given, was solved in \cite{Sch91} in the case where the compact $K$ is semi-algebraic (i.e.~defined by polynomial inequalities). For the $K$-tms problem (and for $K$ semi-algebraic), Curto and Fialkow \cite{CurFia05} obtained a necessary and sufficient condition for a tms to admit a representing measure (see Theorem 1 below). In \cite{HeltonNie2012}, a semidefinite algorithm  was introduced, allowing one to find a representing measure (if it exists), and later generalized to situations where only a subset of moments up to a certain order are known \cite{Nie14}. This algorithm was also used in \cite{NieZhang16} to test positivity of linear maps and separability of matrices in relation with the entanglement problem. More detail on the history of the tms problem can be found in the review \cite{LaurentReview2010}.

The goal of the present paper is to show how the separability problem
for an arbitrary quantum state can be mapped to the $K$-tms problem,
and to use results from the tms literature to elucidate some aspects
of entanglement detection and characterization of
separability. From an analytical point of view, the mapping allows us
to make use of theorems providing necessary and sufficient
separability conditions. Numerically, semi-definite programming yields
an algorithm to obtain an explicit decomposition of separable
states. 

The idea of using semi-definite programming to test for
entanglement was already proposed in
\cite{DohParSpe02,DoPaSp04,DoPaSp05} by Doherty, Parrilo, and
Spedalieri, and independently 
in \cite{EisHylGuhCur04}. 
In \cite{DoPaSp04} an
algorithm was provided which detects entanglement, but this algorithm
never stops if the state is separable. Conversely, the algorithm
proposed in \cite{HulBru05} detects separable states but does not
certify entanglement. 
The algorithms in \cite{DohParSpe02,DoPaSp04,DoPaSp05}   use
the concept of 
``extensions'', i.e.~states in a larger Hilbert space are considered,
such that their partial trace gives back the original state.  By going
to larger and larger extensions, a hierarchy of semi-definite programs (SDPs)
arises whose infeasability at any stage signals that the original
state $\rho_{AB}$ is
entangled. 
The authors of \cite{DohParSpe02,DoPaSp04,DoPaSp05}  add
the request that the extensions have 
positive partial transpose (i.e.~are ``PPT'') as a necessary
criterion for separability. This additional condition can be
implemented at little extra 
cost in the SDP.  Furthermore, they search in the space of ``$N$
Bose-symmetric 
extensions'', where the extended state $\rho_{AB^N}$ (besides being positive
semi-definite and reproducing $\rho_{AB}=\tr_{B^{N-1}}[\rho_{AB^N}]$) is
invariant under projection onto the symmetric subspace of $B^N$. 
These algorithms were further improved in
\cite{NaOwPl09,TamingEnt11,HaNaWu15,ChJiKrLue14}.

The algorithm we propose here gives a unifying mathematical 
framework that 
also uses semi-definite programming and extensions, but in a somewhat more
abstract way, based on a matrix of moments and a  theorem in the 
theory of moment sequences. It provides an elegant solution of the entanglement
problem, and in particular provides a certificate of separability,
together with an explicit decomposition into product states if the
state is separable. Moreover, it applies to arbitrary quantum state
with arbitrary number of constituents and arbitrary symmetries between
the subparts and easily accomodates missing data, i.e.~incompletely
specified states. 

After setting up the notations, we define the $K$-tms problem
(Section \ref{sec:deftms}), explain the procedures and algorithms
allowing to solve it (Section \ref{sec:solvtmsV2}) and then show
explicit numerical results (Section \ref{sec:numericalresults}). {In
  Section \ref{sec:newsol} we show that, conversely, some solutions of
  the entanglement problems may shed light on a particular tms
  problem.} A discussion of the advantages and novelties of our
treatment compared to previous algorithms is provided in the conclusions.

\section{Entanglement and the truncated moment problem}\label{sec:deftms}

To familiarize the reader with the notations in this paper, we will first consider the case of symmetric states of qubits, since in this case the equations are more compact. After that we will explain the general case in the following subsection. 

\subsection{Symmetric qubit case}
\label{subsec:symmetric}

Multi-qubit pure states which are invariant under any permutation of the qubits are called symmetric pure states. Symmetric states are mixtures of symmetric pure states. Such states are formally equivalent to spin states with spin quantum number $N/2$, where $N$ is the number of qubits. This connection can be made explicit with the Dicke states defined by
\begin{equation}
\label{DickeStates}
\ket{D_N^{(k)}}=\mathcal{N}\sum_\pi \ket{\underbrace{0 \dots 0}_{k} \underbrace{1\dots 1}_{N-k}},
\end{equation}
where $\mathcal{N}$ is a normalization constant and the sum runs over all permutations of the qubits. These states with $k \in \{0,\dots,N\}$ form a basis of the symmetric subspace of the Hilbert space $\mathbb C^{2^N}$ of $N$ qubits. We now introduce a convenient way of representing symmetric states as tensors. For a state $\rho$, let 
\begin{equation}
\label{Tensorrep}
X_{\mu_1\mu_2...\mu_N}=\tr\left\{ \rho \, P^{\dagger}_s \sigma_{\mu_1} \otimes \cdots \otimes \sigma_{\mu_N} P_s\right\},
\end{equation}
with $\sigma_0$ the $2\times 2$ identity matrix, $\sigma_{1},\sigma_2,\sigma_3$ the three 
Pauli matrices, and $P_s$ the projector onto the symmetric subspace spanned by Dicke states \eqref{DickeStates}. Then $\rho$ can be expanded \cite{PRL_Tensor} as 
\begin{equation}
\label{projrho}
\rho=\frac{1}{2^N} X_{\mu_1\mu_2...\mu_N} P^{\dagger}_s \sigma_{\mu_1} \otimes \cdots \otimes \sigma_{\mu_N} P_s
\end{equation}
(with summation over repeated indices). The tensor $X_{\mu_1\mu_2...\mu_N}$ is real and invariant under permutation of indices, and verifies
\begin{equation}
X_{0...0}=\tr\rho=1.
\end{equation}
In this representation, the tensor associated with a pure separable symmetric state $\ket{\psi_{sep}}$ of $N$ qubits takes the particularly simple form
\begin{equation}
X_{\mu_1\mu_2...\mu_N}=n_{\mu_1} \cdots n_{\mu_N}
\end{equation}
with $n_0=1$ and $\vec{n}=(n_1,n_2,n_3)$ the Bloch vector of the individual
qubit, $n_1^2+n_2^2+n_3^2=1$. Note that since the state is invariant under the exchange of
qubits, a pure state can only be the tensor product of identical
qubits (with same Bloch vector $\vec{n}$), and a separable pure symmetric state has to be fully
separable \cite{eta_Paper}. 
As a consequence, a symmetric state is separable if and only if its tensor representation can be written as 
\begin{equation}
\label{tensorseparablestatesDef}
X_{\mu_1\mu_2...\mu_N}=\sum_j w_j n_{\mu_1}^{(j)} \cdots n_{\mu_N}^{(j)},
\end{equation}
with $w_j \geqslant 0$, $n_0^{(j)}=1$ and each Bloch vector $\vec{n}^{(j)}$ normalized to 1. This can be equivalently written in an integral form as 
\begin{equation}
\label{integraldefsym}
X_{\mu_1\mu_2...\mu_N}=\int_K x_{\mu_1} x_{\mu_2} \cdots x_{\mu_N} d\mu(\vec{x}),
\end{equation}
with $K=\{\vec{x} \in \mathbb{R}^3  :x_1^2+x_2^2+x_3^2=1\}$ the unit sphere, $x_0=1$, and $d\mu$ a positive measure on $K$. Indeed, if \eqref{tensorseparablestatesDef} holds then the tensor can be written as in \eqref{integraldefsym} with 
\begin{equation}
d\mu(\vec{x})=\sum_j w_j \delta(\vec{x}-\vec{n}^{(j)}).
\end{equation}
Conversely, since the system is finite-dimensional, Carath\'eodory's theorem implies that the integral in \eqref{integraldefsym} can always be reduced to a finite sum as in \eqref{tensorseparablestatesDef}, so that the positive measure can always be expressed as a sum of delta functions. Expressing Eq.~\eqref{integraldefsym} in words, a symmetric state is separable if and only if there exist a positive measure $d\mu$ such that all entries of the tensor $X_{\mu_1\mu_2...\mu_N}$ (for all $\mu_j$, $1 \leqslant j \leqslant N$ and $0\leq\mu_i\leq 3$) are given by moments of that measure.

In order to prepare for the generalization to arbitrary states in the next subsection, let us introduce a more compact notation for Eq.~\eqref{integraldefsym}. For any $N$-tuple $(\mu_1,\ldots,\mu_N)$ we define a triplet $\alpha=(\alpha_1,\alpha_2,\alpha_3)$ of integers such that 
\begin{equation}
\label{xmualpha1}
x_{\mu_1} x_{\mu_2} \cdots x_{\mu_N}=x^\alpha,
\end{equation}
where we use the notation $x^\alpha=x_1^{\alpha_1} x_2^{\alpha_2} x_3^{\alpha_3}$. So e.g.~for $\alpha=(1,3,0)$ we have $x^\alpha=x_1 x_2^3$. The degree of the monomial $x^\alpha$ is denoted $|\alpha|\equiv\sum_i\alpha_i$. We also denote the $X_{\mu_1\mu_2...\mu_N}$ by $y_{\alpha}$, where $\alpha$ corresponds to $(\mu_1\mu_2...\mu_N)$ via Eq.~\eqref{xmualpha1}, so that e.g.~for $N=6$, $y_{(2,1,0)}=X_{000112}$. With this notation we can rewrite \eqref{integraldefsym} as
\begin{equation}
\label{defIntergalSymCompact}
y_\alpha=\int_K x^\alpha d\mu(\vec{x}).
\end{equation}

To test if a symmetric state is separable, a necessary and sufficient condition is therefore that a positive measure  $d\mu$ exists that fulfills \eqref{defIntergalSymCompact} for { all $|\alpha|\leqslant N $.}  Problems of this type are known as \textit{truncated $K$-moment sequence problems} (or $K$-tms problems), and they can be solved by a semi-definite relaxation procedure. Before we describe this method in Section \ref{sec:solvtmsV2} we generalize the description to arbitrary states of finite-dimensional systems. 

\subsection{General case}
\label{subsec:general}

Consider a multipartite quantum state $\rho$ acting on the tensor
product $\mathcal{H}=\mathcal{H}^{(1)} \otimes\mathcal{H}^{(2)}
\otimes \cdots \otimes \mathcal{H}^{(d)}$ of Hilbert spaces
$\mathcal{H}^{(i)}$. For each $i$, let $S_\mu^{(i)},0\leq \mu\leq t_i$ be a set of
$t_i+1$ Hermitian matrices forming an orthogonal basis (with respect to the scalar product tr$A^\dagger B$) of the 
set of bounded linear operators on $\mathcal{H}^{(i)}$, with the
choice that $S_0^{(i)}$ is the identity matrix. An orthogonal basis
of $\mathcal{H}$ is then given by matrices 
\begin{equation}
S_{\mu_1 \mu_2\dots \mu_d}=S^{(1)}_{\mu_1} \otimes S^{(2)}_{\mu_2} \otimes\cdots  \otimes S^{(d)}_{\mu_d}
\end{equation}
and any state can be written as
\begin{equation}
\label{exprho}
\rho=\mathcal{N}X_{\mu_1\mu_2\ldots \mu_d} S_{\mu_1 \mu_2\dots \mu_d}
\end{equation}
where summation over repeated indices is understood, and the normalization constant $\mathcal{N}=\prod_{i=1}^d\mathcal{N}_i$, with $\mathcal{N}_i=1/\sqrt{t_i+1}$, is chosen so that $X_{0\ldots 0}=1$. A quantum state $\rho_{sep}$ is said to be separable (over that particular factorization of $\mathcal{H}$) if it can be written as
\begin{equation}
\label{defrhosep}
\rho_{sep}=\sum_j w_j\, \rho_j^{(1)} \otimes \rho_j^{(2)} \otimes \dots\otimes  \rho_j^{(d)}
\end{equation}
with $w_j \geqslant 0$, and $\rho_j^{(i)}$ density matrices acting on $\mathcal{H}^{(i)}$ \cite{Werner89}. Any $\rho^{(i)}$ acting on $\mathcal{H}^{(i)}$ can be expanded as $\rho^{(i)}=\mathcal{N}_i\sum_{\mu_i} y_{\mu_i}^{(i)}S_{\mu_i}^{(i)}$, with $y^{(i)}$ a real $(t_i+1)$-dimensional vector. The condition $\tr\rho^{(i)}=1$, together with the choice that $S_0^{(i)}$ is the identity matrix and the normalization, implies that $y_0^{(i)}=1$. 

Rewriting condition \eqref{defrhosep} in terms of average values, we get that a state is fully separable if and only if all averaged basis operators can be expressed as
\begin{equation}
\label{matrixS}
\langle S_{\mu_1 \mu_2\dots \mu_d} \rangle_{\rho}=\sum_j w_j \langle S_{\mu_1}^{(1)} \rangle_{\rho_j^{(1)}}\langle S_{\mu_2}^{(2)} \rangle_{\rho_j^{(2)}} \cdots \langle S_{\mu_p}^{(d)} \rangle_{\rho_j^{(d)}}
\end{equation}
with $w_j \geqslant0$, i.e.~the expectation values of all $S_{\mu_1  \mu_2\dots \mu_d}$ are convex combinations of the product of local expectation values. This condition can be reexpressed in terms of the coefficients $X_{\mu_1\mu_2\ldots \mu_d}$ of $\rho_{sep}$ in the expansion \eqref{exprho} and the coefficients $y_{a_i}^{(i;j)}$, $1\leq a_i\leq t_i$, in the expansion $\rho_j^{(i)}=\mathcal{N}_i\sum_{\mu_i} y_{\mu_i}^{(i;j)} S_{\mu_i}^{(i)}$, with $y_0^{(i;j)}=1$. Separability is then equivalent to the existence of $w_j \geqslant0$ and real numbers $y_{a_i}^{(i;j)}$, $1\leq a_i\leq t_i$, such that for all $\mu_i$ with $0\leq \mu_i\leq t_i$ one has
\begin{equation}
\label{ysumyk}
X_{\mu_1\mu_2\ldots \mu_d}=\sum_j w_j\, y_{\mu_1}^{(1;j)} y_{\mu_2}^{(2;j)} \cdots y_{\mu_d}^{(d;j)}
\end{equation}
and $\sum_{\mu_i}y_{\mu_i}^{(i;j)} S_{\mu_i}^{(i)}\geqslant 0$ for all
$i$ and $j$. This latter condition comes from the fact that each
$\rho_j^{(i)}=\mathcal{N}_i \sum_{\mu_i} y_{\mu_i}^{(i;j)}
S_{\mu_i}^{(i)}$ appearing in \eqref{defrhosep} is a density matrix,
and thus has to be positive. Since matrices are Hermitian and thus
have all their eigenvalues real, one can use Descartes sign rule to
express this positivity condition as inequalities on the coefficients
of the characteristic polynomial of $\rho_j^{(i)}$. Each of these
coefficient is a linear combination of traces of powers of
$\rho_j^{(i)}$, and therefore a polynomial in the variables
$y_{\mu}^{(i;j)}$. Thus, each vector ${\bf
  y}^{(i;j)}=(y_1^{(i;j)},\ldots,y_{t_i}^{(i;j)})$ is restricted to a
certain compact subset $K^{(i)}\subset \mathbb{R}^{t_i}$ defined by
some polynomial inequalities, e.g.~for a qubit the polynomial is a
  quadratic equation of the Bloch vector, restricting its maximal
  length to one. Defining the compact $K=K^{(1)} \times
K^{(2)} \times \cdots \times K^{(d)}\subset\mathbb{R}^{n}$,
$n=\sum_it_i$, and the vector
$\vec{y}^{(j)}=(\vec{y}^{(1;j)},\vec{y}^{(2;j)},\dots,\vec{y}^{(d;j)})\in\mathbb{R}^{n}$,
the positivity condition on the partial density matrices amounts to
impose that $\vec{y}^{(j)}\in K$ with $K$ a compact defined by
polynomial inequalities. Equation \eqref{ysumyk} can then be rewritten, for
$0\leq \mu_i\leq t_i$, as 
\begin{equation}
\label{integralform}
X_{\mu_1\mu_2\dots \mu_d}=\int_K x_{\mu_1}^{(1)} x_{\mu_2}^{(2)} \cdots x_{\mu_d}^{(d)} d \mu(\vec{x})
\end{equation}
with $x_0^{(i)}=1$, $\vec{x}=\left(\vec{x}^{(1)},\vec{x}^{(2)},\dots,\vec{x}^{(d)}\right)\in\mathbb{R}^{n}$, $\vec{x}^{(i)}=(x^{(i)}_{a})_{1\leq a\leq t_i}\in\mathbb{R}^{t_i}$, and $d \mu$ the measure over $\mathbb{R}^{n}$ defined by
\begin{equation}
\label{defdmu}
d \mu(\vec{x})=\sum_j w_j\,\delta(\vec{x}-\vec{y}^{(j)}).
\end{equation}
Equation \eqref{integralform} is the generalization of the symmetric case Eq.~\eqref{integraldefsym}, the difference being that each Hilbert space $\mathcal{H}^{(i)}$ has its own set of variables $(x^{(i)}_{a})_{1\leq a\leq t_i}$. As in the symmetric case, the existence of an arbitrary measure $d \mu(\vec{x})$ such that \eqref{integralform} holds is equivalent to the existence of a 'discrete' measure of the form \eqref{defdmu}, since one can apply Carath\'eodory's theorem to our finite dimensional Hilbert spaces. The separability problem, for a state given by \eqref{exprho}, is thus equivalent to the question whether a positive measure $d\mu$ with support $K$ exists whose moments coincide with the coordinates $X_{\mu_1\mu_2\dots\mu_d}$ of the state. 

We now rewrite Eq.~\eqref{integralform} in a more compact form. Let us relabel the entries of $\vec{x}$ as $\vec{x}=\left(x_1,x_2,\dots,x_n\right)$, and introduce the notation $x^\alpha\equiv\prod_{i=1}^n x_i^{\alpha_i}$, where $\alpha=(\alpha_1,\alpha_2,\dots,\alpha_n)$ is a vector of integers. For instance for two qubits we have $\vec{x}=\left(x^{(1)}_1,x^{(1)}_2,x^{(1)}_3,x^{(2)}_1,x^{(2)}_2,x^{(2)}_3\right)=(x_1,x_2,\dots,x_6)$. For any given tuple $(\mu_1,\dots,\mu_d)$, there exists an index $\alpha$ such that 
\begin{equation}
\label{mapmualpha}
x_{\mu_1}^{(1)} x_{\mu_2}^{(2)} \cdots x_{\mu_d}^{(d)}=x^\alpha.
\end{equation}
Thus, $\alpha_1$ counts the number of $x^{(1)}_1$ in the monomial $x_{\mu_1}^{(1)} x_{\mu_2}^{(2)} \cdots x_{\mu_d}^{(d)}$, $\alpha_2$ counts the number of $x^{(1)}_2$, and so on until $\alpha_n$, which counts the number of $x^{(d)}_{t_d}$. For instance for a bipartite state of $d=2$ qubits,  $(\mu_1,\mu_2)=(2,3)$ corresponds to $\alpha=(0,1,0,0,0,1)$ or to the monomial $x_2^{(1)} x_3^{(2)}$, while $(\mu_1,\mu_2)=(1,0)$ corresponds to $\alpha=(1,0,0,0,0,0)$ or to the monomial $x_1^{(1)}$. As each monomial $x_{\mu_1}^{(1)} x_{\mu_2}^{(2)} \cdots x_{\mu_d}^{(d)}$ contains at most one variable of each type $x^{(i)}$, the vector $\alpha$ is such that each tuple $(\alpha_1,\ldots, \alpha_{t_1})$, $(\alpha_{t_1+1},\ldots, \alpha_{t_1+t_2})$,$\ldots$, contains at most one 1. For instance for qubits, where $t_i=3$, each triplet $(\alpha_{3i+1},\alpha_{3i+2},\alpha_{3i+3})$ must therefore contain at most one 1. 

If we denote $X_{\mu_1\mu_2\dots \mu_d}$ by $y_\alpha$, where $\alpha$ is the index corresponding to the tuple $(\mu_1,\dots,\mu_d)$ via \eqref{mapmualpha}, then Eq.~\eqref{integralform} can be simply rewritten as
\begin{equation}
\label{integralform2}
y_\alpha=\int_K x^\alpha d \mu(\vec{x}).
\end{equation}
Hence, a state is separable if and only if all its coordinates $y_\alpha$ can be written as in Eq.~\eqref{integralform2}, with $d\mu$ a positive measure.

\subsection{Examples and special cases}
\label{subsec:Different_Partitions}
The general setting of the previous Subsection allows one to test separability for a given fixed partition. For example, in order to check full separability for a three-qubit state $\rho$ one has to consider $d=3$ sets of variables $x^{(i)}_{a_i}$, each set being associated with a qubit (thus with $1\leq a_i\leq 3$), and $K$ is the product of three Bloch spheres. One then relabels the coordinates $X_{\mu_1\mu_2\mu_3}$ of $\rho$ as $y_\alpha$ and the variables as $(x_1,\ldots,x_n)$ with $n=9$, in order to get Eq.~\eqref{integralform2}. Among all 9-tuples $\alpha=(\alpha_1,\ldots,\alpha_n)$, only the 64 values which correspond to some triplet $(\mu_1,\mu_2,\mu_3)$ for $0\leq \mu_i\leq 3$ via \eqref{mapmualpha} have to be considered. The state $\rho$ is separable if and only if there exists a measure $d\mu$ such that Eq.~\eqref{integralform2} is fulfilled for all these $\alpha$.

 However if one is only interested in the question of entanglement of the first two qubits with respect to the third one, one would have to take the first two qubits as a 4-level system. There would then be two sets of variables in Eq.~\eqref{integralform}, the first one with $t_1=15$ variables (characterizing the density matrix of a 4-level system), and the second with $t_2=3$ variables (characterizing a mixed qubit state). Thus one has $d=2$ and $n=18$ variables. Finding whether or not \eqref{integralform2} can be solved answers the question whether or not the third qubit is entangled with the first two, while ignoring any entanglement between the first two qubits.\\

It is instructive to see how the symmetric case of Subsection \ref{subsec:symmetric} can be recovered from the general case. As we saw in Subsection \ref{subsec:symmetric}, the problem of finding whether a symmetric $N$-qubit state is fully separable can be cast into the form \eqref{defIntergalSymCompact}, with $K$ the 2-sphere and $\alpha$ running over triplets of integers with $|\alpha|\leq N$. Applying the general case to the $N$-qubit case implies $d=N$ parties, and the Hilbert space $\mathcal{H}$ is decomposed as $\mathcal{H}=\mathcal{H}^{(1)} \otimes\mathcal{H}^{(2)}\otimes \cdots \otimes \mathcal{H}^{(N)}$. Each Hilbert space $\mathcal{H}^{(i)}$ has its own set of variables $x^{(i)}_{a_i}$, $1\leq a_i\leq 3$, appearing in the right-hand side of Eq.~\eqref{integralform}. The basis $S_\mu^{(i)}$ in the decomposition $\rho^{(i)}=\frac12 \sum_\mu y_{\mu}^{(i)}S_\mu^{(i)}$ is the Pauli basis, the vectors $(y^{(i)}_a)_{1\leq a\leq 3}$ are the Bloch vectors and the compact $K^{(i)}$ such that $\sum_\mu y_{\mu}^{(i)}S_\mu^{(i)}$ is positive is the Bloch ball. Symmetry then implies that the variables corresponding to each Hilbert space are not independent but equal, so that one has to require $x_\mu^{(i)}=x_\mu$ for all $i$ and $\mu$, and replace the compact $K^{(1)}\times K^{(2)} \times \cdots \times K^{(d)}$ by $K=K^{(1)}$, a single Bloch sphere. To account for the fact that the different sets of variables should no longer be distinguished, the $n$-tuple $\alpha$ in \eqref{integralform2} should be replaced by the triplet $(\sum_i\alpha_{3i+1},\sum_i\alpha_{3i+2},\sum_i\alpha_{3i+3})$ giving the multiplicities of $x_1,x_2,x_3$. The entries of the triplet can now take values larger than one. Since Eq.~\eqref{integralform2} and Eq.~\eqref{defIntergalSymCompact} coincide, the symmetric and the general case are in essence the same problem; the difference between them lies only in the definition of the compact $K$ supporting the measure, and also in the set of tuples $\alpha$ considered. 
 
The general formalism \eqref{integralform2} allows us in fact to play with any kind of constraint, just by adjusting the sets of variables and $\alpha$ vectors accordingly. The symmetric case explained above is just one example, but this method is general. For instance if one wants to impose a symmetry between two of the subsystems one just has to equate the sets of independent variables. This adjustment can be easily generalized to test for entanglement for any type of partition. The algorithms for the truncated moment problem that we will present in Section \ref{sec:solvtmsV2} provide a solution to all these cases.

\subsection{Partial knowledge of a state}
\label{subsec:partialknowledgeofthestate}

An interesting question in practical application is whether or not a
partial set of measurement results is compatible with a separable
state. If for example a state tomography is not carried to its end, or
if only local measurements are available, can one in some instances
infer that the state was entangled? Another
interesting question is whether the partial traces of a state can be
used to show entanglement of the global state even if all the {reduced
  states} are separable \cite{GuehneEmergentEnt}. 

Such problems of partial knowledge can be formulated in the form of
Eq.~\eqref{integralform2} very simply. The only change is the range of
tuples over which $\alpha$ varies: since the unknown measurements
correspond to unknown $y_\alpha$, these values of $\alpha$ should not
be taken into account as constraints on $d\mu$. If for example 
only the results of local measurements are known, only averages of the form $\langle S^{(1)}_0 \otimes S^{(2)}_0 \otimes\cdots S^{(i)}_\mu\cdots \otimes S^{(d)}_0\rangle $ are known (recall that $S^{(i)}_0$ is the identity matrix). Therefore one only knows the values of $y_\alpha$ such that the $\alpha=(\alpha_1,\dots,\alpha_n)$ have only one non-zero entry. This problem can then be solved in the same way as the general one, just by putting no constraint on the unknown moments.

\section{TMS Problems: definitions and solutions}
\label{sec:solvtmsV2}

Identifying the entanglement problem with the $K$-tms problem allows us to use analytical results and numerical methods from the tms literature to get insight in entanglement theory. We now introduce the mathematical formalism used to describe and solve the tms problem.

\subsection{Truncated moment problems}
\label{subsec:tms}

A truncated moment sequence (tms) of degree $d$ is a finite set of numbers $y=(y_\alpha)_{|\alpha|\leqslant d}$ indexed by $n$-tuples $\alpha=(\alpha_1,\ldots,\alpha_n)$ of integers $\alpha_i\geqslant 0$ such that $|\alpha|=\sum_i\alpha_i\leqslant d$ \cite{Nie14}. The truncated $K$-moment problem consists in finding conditions under which there exists a (positive) measure $d\mu$ such that each moment $y_\alpha$ with $|\alpha|\leqslant d$ can be represented as an integral of the form
\begin{equation}
\label{ktms_eq}
y_\alpha=\int_{K}x^{\alpha} d\mu(\vec{x})
\end{equation}
with $\vec{x}=(x_1,\ldots,x_n)\in\mathbb{R}^n$, $x^\alpha=x_1^{\alpha_1} x_2^{\alpha_2} \ldots x_n^{\alpha_n}$, and $d\mu$ a measure supported on a semi-algebraic set 
\begin{equation}
\label{defK}
K=\{\vec{x} \in \mathbb{R}^n| g_1(\vec{x})\geqslant0,\cdots,g_m(\vec{x})\geqslant0 \}
\end{equation}
with $g_i(\vec{x})$ multivariate polynomials in the variables $x_1,\ldots,x_n$. If such a measure exists it can be written as the sum of  delta functions
\begin{equation}
\label{tms delta function}
d\mu(\vec{x}) =\sum _{j=1}^r w_j \delta\left(\vec{x}-\vec{y}^{(j)}\right)
\end{equation}
with some finite $r$, $w_j> 0$ and $\vec{y}^{(j)}\in K$. Such a measure is then called a \textit{finitely atomic representing measure}. Equation \eqref{ktms_eq} is nothing but Eq.~\eqref{defIntergalSymCompact}, where $K$ is the Bloch sphere, $d=N$, and $n=3$. Therefore the entanglement problem for symmetric states is a special case of $K$-tms problem.\\

The $\mathcal{A}K$-tms problem \cite{Nie14} is a generalization of the $K$-tms problem in which moments $y_\alpha$ are known only for a finite subset $\mathcal{A}\subset \mathbb{N}^n$ of indices of degree $|\alpha|\leqslant d$. The only difference with the $K$-tms problem is that Eq.~\eqref{ktms_eq} now has to be fulfilled only for $\alpha\in\mathcal{A}$. This is exactly the situation found in the general case of Subsection \ref{subsec:general}. Indeed, in that case, we showed that $K$ is defined by polynomial inequalities, so that it is a semi-algebraic compact set. Moreover, only indices $\alpha$ associated with some tuple $(\mu_1,\dots,\mu_d)$ for $0\leqslant \mu_i\leqslant t_i$ do correspond to a certain moment $y_\alpha$, so that a restriction on indices $\alpha$ is required. This is also the situation encountered in Subsection \ref{subsec:partialknowledgeofthestate}, where the state is only known partially. All these cases therefore correspond to the $\mathcal{A}K$-tms problem, and can in fact be solved in the same way as the $K$-tms problem, only with fewer constraints (since less moments are fixed).\\

In all what follows, to ease notations, we will only treat the original $K$-tms problem where all moments $y_\alpha$ with $|\alpha|\leqslant d$ are known. However, we must stress that the $\mathcal{A}K$-tms problem is treated in exactly the same way, just by considering $\alpha\in\mathcal{A}$ rather than $|\alpha|\leqslant d$ in all equations involving that restriction.

\subsection{Moment matrices}
\label{subsec:maths}
Let us now present the mathematical setting for the $K$-tms problem
defined by Eq.~\eqref{ktms_eq}. Let $y=(y_\alpha)_{|\alpha|\leqslant
  d}$ be a tms of degree $d$, with $\alpha=(\alpha_1,\ldots,\alpha_n)$
being $n$-tuples of integers. The integrand                                  
in the right-hand side of Eq.~\eqref{ktms_eq} is a 
monomial in $n$
variables $(x_1,\ldots,x_n)$ of degree less than $d$. Any 
polynomial of degree less than $d$ can be written as a vector in the basis of monomials ordered in degree-lexicographic order (that is, monomials are sorted by order and within each order in a lexicographic order). For instance for $n=3$ and $d=2$ the monomial basis is $\{1,x_1,x_2,x_3,x_1^2,x_1x_2,x_1x_3,x_2^2,x_2x_3,x_3^2\}$, and a polynomial such as e.g.~$p(\vec{x})=7x_3-3x_2^2+2$ would be written as the vector $(2,0,0,7,0,0,0,-3,0,0)$. The components of the vector representing $p(\vec{x})$ are coefficients $p_\alpha$ such that $p(\vec{x})=\sum_\alpha p_\alpha x^\alpha$.

For any integer $k\leqslant d/2$, let $M_k(y)$ be the matrix defined by
\begin{equation}
\label{def_M}
M_k(y)_{\alpha\beta}=y_{\alpha+\beta},\qquad |\alpha|,|\beta| \leqslant k. 
\end{equation}
It is called the {\it moment matrix} of order $k$ associated with the tms $y$. A necessary condition for a tms to admit a representing measure as in \eqref{ktms_eq} is that the moment matrix of any order is positive-semidefinite. Indeed, if \eqref{ktms_eq} holds, then for any vector $p=(p_\alpha)_{|\alpha|\leqslant k}$ representing a polynomial $p(\vec{x})$ of degree $k$ or less we have
\begin{align} \nonumber \label{MomentMatrixPositive}
p^T M_k(y) p=\sum_{|\alpha|,|\beta| \leqslant k} p_\alpha y_{\alpha+\beta} p_\beta=\hfill\\
\hfill\sum_{|\alpha|,|\beta|\leqslant k} p_{\alpha}  p_{\beta} \int_K x^{\alpha+\beta} d\mu(\vec{x}) =\int_K p(\vec{x})^2 d\mu(\vec{x}) \geqslant 0,
\end{align}
so that $M_k(y)$ a is positive-semidefinite matrix \cite{CurFia05, HeltonNie2012}. 

Other necessary conditions can be obtained from the polynomial constraints $g_i(\vec{x})\geqslant0$ which define the set $K$ in \eqref{defK}. For any polynomial $g$ of degree $\text{deg}(g)\geqslant 1$, one can define a 'shifted tms' of degree $d-\text{deg}(g)$ as
\begin{equation}
(g\star y)_\alpha=\hspace{-.3cm}\sum_{|\gamma|\leqslant deg(g)}\hspace{-.3cm} g_{\gamma} y_{\alpha+\gamma},\quad |\alpha|\leqslant d-\text{deg}(g).
\end{equation}
Let $d_g=\lceil\text{deg}(g)/2\rceil$ (we denote by $\lceil x \rceil$ the smallest integer larger than $x$ and by $\lfloor x \rfloor$ the largest integer smaller than $x$). Applying definition \eqref{def_M}, one can define the $(k-d_g)$th moment matrix of $g\star y$, for any integer $k$ such that $0\leqslant k-d_g\leqslant [d-\text{deg}(g)]/2$, by $M_{k-d_g}( g\star y)_{\alpha\beta}=(g\star y)_{\alpha+\beta}$. This matrix is called the $k$th-order {\it localizing matrix} of $g$ \cite{HeltonNie2012}. In explicit form, it reads
\begin{align}
\label{ShiftetMomentMatrixDef} 
M_{k-d_{g}}(g\star y)_{\alpha\beta}=\hspace{-.3cm}\sum_{|\gamma|\leqslant deg(g)}\hspace{-.3cm} g_{\gamma} y_{\alpha+\beta+\gamma} ,\quad &|\alpha|,|\beta| \leqslant k-d_{g}. 
\end{align}
Using the fact that $\lfloor(d-\text{deg}(g))/2 \rfloor=\lfloor d/2\rfloor-d_g$, we have that the $k$th-order localizing matrix is defined for any integer $k$ such that $d_g \leqslant k\leqslant d/2$ (the definition of $d_g$ has been precisely chosen in such a way that the upper bound $k\leqslant d/2$ is the same as that for the $k$th-order moment matrix). If a tms admits a representing measure then any $k$th order localizing matrix is necessarily positive-semidefinite: indeed for any vector $p=(p_\alpha)_{|\alpha|\leqslant k-d_{g}}$ representing a polynomial $p(\vec{x})$ with degree $k-d_{g}$ or less we have
\begin{align}
\label{ShiftetMomentPositive} \nonumber
p^T M_{k-d_{g}}(g\star y) p&=\hspace{-.3cm}\sum_{|\alpha|,|\beta|\leqslant k-d_{g}}\hspace{-.3cm} p_\alpha p_\beta \sum_{|\gamma|\leqslant deg(g)} g_\gamma  y_{\alpha+\beta+\gamma}\\ 
&=\int_K g(\vec{x}) p(\vec{x})^2 d\mu(\vec{x}) \geqslant 0,
\end{align} 
which is positive because $g$ is positive on $K$ by the definition \eqref{defK}. Another way of seing that is to observe that if $y$ admits a positive representing measure then so does the shifted tms $g\star y$.

As moment matrices of order $k'$ are submatrices of matrices of order $k$ if $k'\leqslant k$ it suffices to consider the largest possible value for $k$ to get the strongest necessary conditions. For a tms $y$ of order $d$, the above analysis leads to the necessary condition $M_{\lfloor d/2\rfloor}(y)\geqslant 0$. If the compact $K$ is defined as in \eqref{defK} by polynomial inequalities, the localizing matrices for each polynomial $g_i$, $1\leq i \leq m$, have to be positive, namely $M_{\lfloor d/2\rfloor-d_{g_i}}(g_i\star y)\geqslant 0$, $d_{g_i}=\lceil\text{deg}(g_i)/2\rceil$.

\subsection{A necessary and sufficient condition}
\label{subsec:CNS}
The above conditions are only necessary conditions. A sufficient condition was obtained in \cite{CurFia05} for even-order tms. We formulate it following Theorem 1.1 of \cite{HeltonNie2012}. Namely, if a tms $z$ of even order $2k$ is such that its $k$th order moment matrix and all $k$th order localizing matrices are positive, and if additionally
\begin{equation}
\label{flatcond}
\mbox{rank} M_k(z)=\mbox{rank} M_{k-d_0}(z)
\end{equation}
with $d_0=\max_{1\leq i \leq m}\{1,\lceil\mbox{deg}(g_i)/2\rceil\}$, then the tms $z$ admits a representing measure composed of $r=\mbox{rank} M_k(z)$ delta functions. Note that the rank condition already appeared in \cite{NaOwPl09} under the name rank-loop, using a result from \cite{HoLeViCi00}. 

As the above condition is only sufficient, a tms $y$ admitting a
representing measure does not necessarily fulfill
\eqref{flatcond}. However, one can search for an extension $z$ of $y$
which fulfills it. An extension of a tms $y$ of degree $d$ is defined
as any tms $z$ of degree $2k$ with $2k>d$, such that
$z_\alpha=y_\alpha$ for all $|\alpha|\leqslant d$. A extension $z$ is
called flat if it satisfies Eq.~\eqref{flatcond}. If $z$ verifies the
sufficient conditions above, then it has a representing measure, and
so does $y$ as a restriction of $z$. This allows us to formulate the
following necessary and sufficient condition for the existence of a
representing measure. 
\begin{theorem}[\cite{CurFia05} (see also Theorem 1.2 of
  \cite{HeltonNie2012})]\label{th1} 
A tms $(y_\alpha)_{|\alpha|\leqslant d}$ admits a representing measure
supported by $K$ if and only if there exists a flat extension
$(z_\beta)_{|\beta|\leqslant 2k}$ with $2k>d$ such that $M_{k}(z) \geq
0$, and $M_{k-d_{g_i}}(g_i\star z) \geqslant 0$ for $i=1,\ldots, m$. 
\end{theorem}
This theorem can be implemented as a semi-definite program, as shown in
Section \ref{sec:solvtms}. It has been extended to an abritrary
$\mathcal A$K-tms in proposition 3.3 in \cite{Nie14}. With the
identifications made in 
Sec.\ref{sec:deftms} between the entanglement and the tms problem,
these results can be reformulated as a necessary   and sufficient  
condition for separability of an arbitrary quantum state: 
\begin{theorem}
A state $\rho$ is separable if and only if its coordinates
$X_{\mu_1\mu_2\dots \mu_d}$ defined in \eqref{exprho} correspond to a
tms $(y_\alpha)_{\alpha \in \mathcal A}$ such that there exists a flat
extension  
$(z_\beta)_{|\beta|\leqslant 2k}$ with $2k> d$, $M_{k}(z)
\geqslant 0$, and $M_{k-d_{g_i}}(g_i\star z) \geqslant 0$ for
$i=1,\ldots, m$. 
\end{theorem}

\subsection{Semi-definite program and the entanglement problem}
\label{sec:solvtms}

For a given tms $(y_\alpha)_{|\alpha|\leqslant d}$, finding a
extension $(z_\beta)_{|\beta|\leqslant 2k}$ as in the theorem above
amounts to constructing a positive matrix
$M_k(z)_{\alpha\beta}=z_{\alpha+\beta}$ with some entries given,
namely $z_\alpha=y_\alpha$ for $|\alpha|\leqslant d$, and 
constraints of positivity of moment matrices and localizing matrices,
which are linear in the $z_\alpha$. This type of problem corresponds
to what is known in numerical analysis as semi-definite program (SDP)
problems. Here, the variables of the SDP are the $z_\beta$ for
$|\beta|\leqslant 2k$. The smallest extension order is $k_0=\lfloor
d/2\rfloor+1$. All the constraints of Theorem 1 can be directly
implemented in the SDP apart from the flatness condition
\eqref{flatcond}.  If also the flatness condition could be implemented
efficiently then 
$P=NP$ \cite{LaurentReview2010}. To take into account the flatness
condition, the idea \cite{Nie14} 
is to consider the SDP 
\begin{align}
\label{SDPGeneral}
\min_z \sum_{\alpha,|\alpha|\leqslant k_0}  R_\alpha z_\alpha \quad \mbox{such that} \\ 
\label{SDPGeneral1}
M_k(z)\geqslant0\\
\label{SDPGeneral2}
M_{k-d_i}(g_i\star z)\geqslant 0\quad\textrm{for }i=1,\ldots,m\\ 
\label{SDPGeneral3}
z_{\alpha}=y_{\alpha} \,\,\mbox{for} \,  |\alpha|\leqslant d.
\end{align}
The coefficients $R_\alpha$ are chosen randomly, but in order to
ensure that $\sum_{\alpha,|\alpha|\leqslant k_0}  R_\alpha z_\alpha $
has indeed a global minimum, the polynomial  $R(\vec{x})=\sum_\alpha R_\alpha x^\alpha$ is taken as a sum-of-squares polynomial of degree $2k_0$. When the order of the extension $k$ is increased, the polynomial is kept the same, so that minimization is realized only on the $z_\beta$ with $|\beta|\leqslant 2k_0$.

\begin{figure}[t!]
\begin{center}
\caption{Flow diagram to visualize the algorithm. ''Solve SDP'' refers to (\ref{SDPGeneral})-(\ref{SDPGeneral3})}
\label{fig:Flow}
\includegraphics[width=0.44\textwidth]{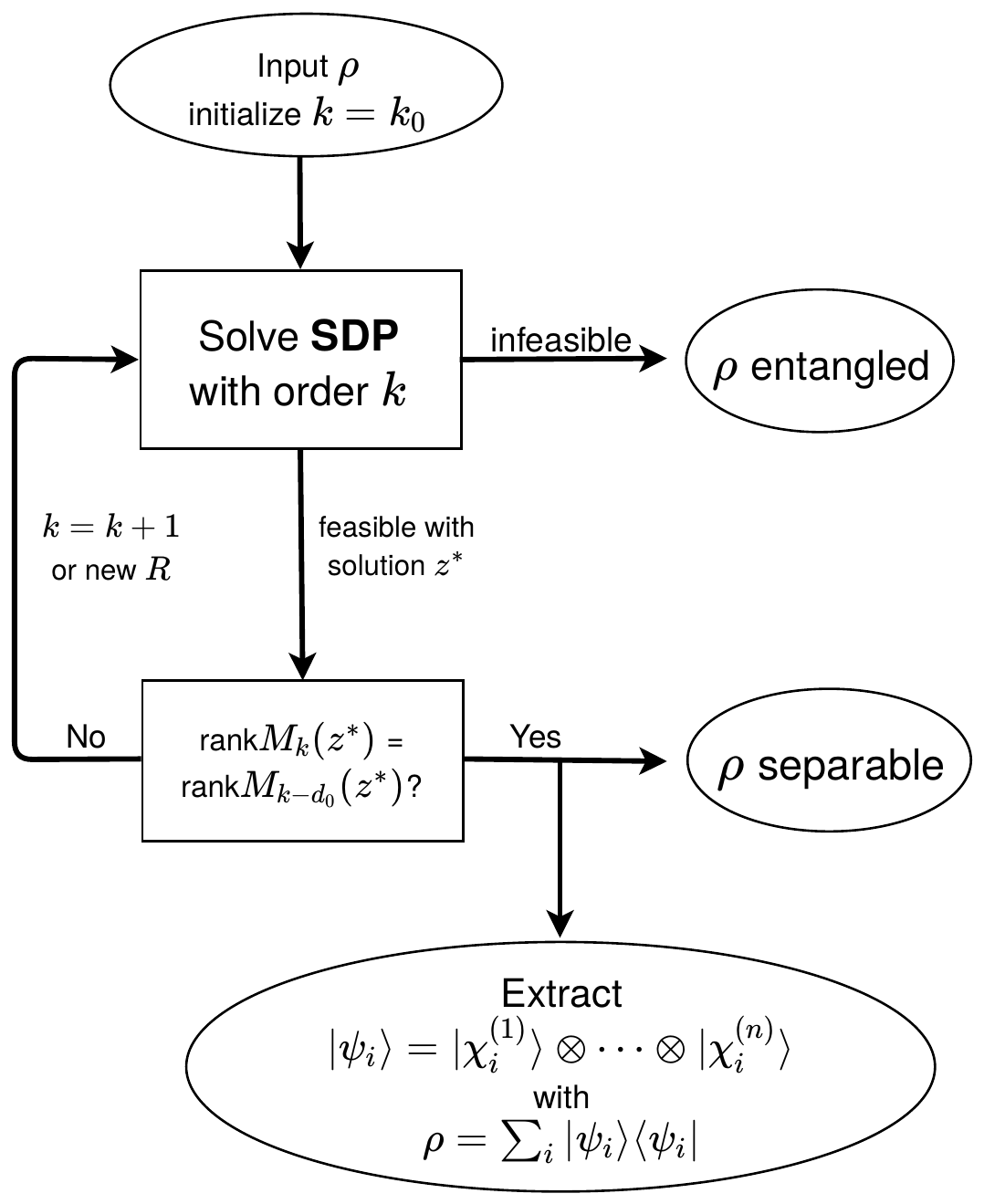}
\end{center}
\end{figure}

According to the theorems above, finding a representing measure, or
finding a decomposition into a mixture of separable product states,
amounts to finding an extension $z$ that fulfills the constraints
\eqref{SDPGeneral1}--\eqref{SDPGeneral3}, i.e.~such that the SDP is
"feasible", and that also fulfills the rank condition
\eqref{flatcond}. We can now propose an algorithm which, for any entangled
state, 
provides a certificate of entanglement, and for a separable state
usually halts at the first iteration $k=k_0$ and provides a
decomposition into pure product states. 
This algorithm is
illustrated in Fig.~\ref{fig:Flow}. One runs the algorithm by starting
from the lowest possible extension order $k=k_0$ and increasing
$k$. If there exists an order $k$ such that the SDP is infeasible,
then the tms $y$ admits no representing measure. In terms of
entanglement, this means that the quantum state whose coordinates are
given by the $y_\alpha$ is entangled. If, on the contrary, the SDP
problem is feasible at some order $k$ (i.e.~if all constraints can be
met) and if for that value of $k$ the extension obtained fulfills
\eqref{flatcond}, then the tms $y$ admits a representing measure, and
the corresponding quantum state  is separable with respect to the
multipartite factorization of Hilbert space considered. The algorithm
remains inconclusive as long as the SDP remains feasible but with an
extension which is not flat. In such a case, one can either repeat the
SDP with the same $k$ and a different $R$, or increase the order $k$
by one. As soon as the rank condition is met, or the SDP becomes
infeasible, the algorithm stops and gives a certificate of
separability, or entanglement. The only situation where the algorithm
does not give an answer in a finite number of steps is the case where
extensions are found for any $k$ and all choices of $R$, but are
never flat. 

When the algorithm stops with
a feasible flat extension $z$ it is possible to extract a representing
measure as a sum of rank[$M_k(z)$] delta functions
\cite{HeltonNie2012}, which provides an explicit factorization of the
separable quantum state.  Indeed, suppose the algorithm stops at order
$k$ and gives an extension $z=z^*$ which optimizes \eqref{SDPGeneral}
and fulfills the rank condition \eqref{flatcond}. If the moment matrix
of the optimal solution $M_k(z^*)$ has 
rank $r$, then it is possible to calculate an explicit decomposition of the form
\begin{equation}
\label{decompmkz}
M_k(z^*)_{\alpha\beta}=\sum_{j=1}^r w_j\, x^*(j)^\alpha x^*(j)^\beta,\quad |\alpha|,|\beta| \leqslant k,
\end{equation} 
with $w_j \geq 0 $,  $\sum_j w_j=1$, and $\vec{x}^*(j)\in K$, with the
methods described in \cite{ExtractingSolutions} and implemented in the
Matlab package Gloptipoly 3 \cite{GloptiPoly3} (see Appendix
\ref{app1}). These $r$ vectors yield $r$ delta functions in the
decomposition of the representing measure, and for a  separable
quantum state they yield an explicit decomposition as a sum of $r$
factorized states.  

\section{Implementation and numerical results}
\label{sec:numericalresults}

\subsection{Two-qubit symmetric states}
\label{AppendixExample}

We now apply this tms approach to some concrete examples of entanglement detection, starting with the simplest case of two-qubit symmetric states. Any state $\rho$ can be expanded as in Eq.~\eqref{projrho} with $N=2$. The tms problem is given by \eqref{defIntergalSymCompact} with $d=N=2$ and $n=3$ variables. We can choose to obtain a decomposition of the state either into mixed states, in which case the compact $K$ should be taken as the unit ball, or into pure states, where $K$ has to be the unit sphere. Here we consider the pure state decomposition, so that we define $K =\{ \vec{x}\in \mathbb{R}^3| g(\vec{x})=0\}$ with $g(\vec{x})=x_1^2 +x_2^2+x_3^2-1$ (equality $g(x)=0$ obviously means that $K$ is semi-algebraic and defined by the two polynomials $g\geqslant0$ and $-g\geqslant0$). The measure $d\mu$ must satisfy constraints such as
\begin{align}
y_{110}=\int_{||\vec{x}||=1} x_1 x_2 d\mu(\vec{x}) \quad \mbox{or} \quad y_{002}=\int_{||\vec{x}||=1} x_3^2d\mu(\vec{x}) \,
\end{align}
where $||\vec{x}||^2=x_1^2+x_2^2+x_3^2$ and the $y_\alpha$ are the entries corresponding to the $X_{\mu_1\mu_2}$. 

The necessary condition given in Subsection \ref{subsec:maths} is positivity of the moment matrix of order $d/2=1$, that is, $M_1(y)\geq 0$, with 
\begin{align}
\label{stateexample}
M_1(y)=\left(
\begin{array}{cccc}
 y_{000} & y_{100} & y_{010} & y_{001} \\
 y_{100} & y_{200} & y_{110} & y_{101} \\
 y_{010} & y_{110} & y_{020} & y_{011} \\
 y_{001} & y_{101} & y_{011} & y_{002} \\
\end{array}
\right).
\end{align}
Solving the entanglement problem in this case amounts to constructing a tms $(z_{\beta})_{|\beta |\leqslant 2k}$ which is a flat extension of $y$. Since $k_0=2$, the lowest-order moment matrix of the extension is $M_2(z)$, which is a $10\times 10$ matrix whose upper left $4\times 4$ block is the matrix \eqref{stateexample}. The conditions of Theorem 2 imply that we look for an extension such that $M_2(z)\geq 0$ and $M_1(g*z)\geq 0$, where
\begin{widetext}
\begin{align}
M_1(g*z)= \left(
\begin{array}{cccc}
 z_{000}-z_{200}-z_{020}-z_{002} & z_{100}-z_{300}-z_{120}-z_{102} & z_{010}-z_{210}-z_{030}-z_{012} & z_{001}-z_{201}-z_{021}-z_{003} \\
 z_{100}-z_{300}-z_{120}-z_{102} & z_{200}-z_{400}-z_{220}-z_{202} & z_{110}-z_{310}-z_{130}-z_{112} & z_{101}-z_{301}-z_{121}-z_{103} \\
 z_{010}-z_{210}-z_{030}-z_{012} & z_{110}-z_{310}-z_{130}-z_{112} & z_{020}-z_{220}-z_{040}-z_{022} & z_{011}-z_{211}-z_{031}-z_{013} \\
 z_{001}-z_{201}-z_{021}-z_{003} & z_{101}-z_{301}-z_{121}-z_{103} & z_{011}-z_{211}-z_{031}-z_{013} & z_{002}-z_{202}-z_{022}-z_{004} \\
\end{array}
\right)
\end{align}
\end{widetext}
is the $4\times 4$ localizing matrix of $z$. The SDP is then to find
$\min_z\sum_\alpha R_\alpha z_\alpha$, with $R$ an arbitrary given
list of coefficients so that $\sum_\alpha R_\alpha z_\alpha$ is positive and bounded, under the constraints that $M_2(z)\geqslant0$, $M_1(g*z)\geqslant0$ and $z_{\alpha}=y_{\alpha}$ for $|\alpha|\leqslant 2$.

The point of this subsection was to illustrate the different
ingredients of our algorithm. In fact, in this case, the necessary
condition $M_1(y)\geq 0$ is necessary and sufficient. Indeed, $M_1(y)$
is exactly the $4\times 4$ matrix $(X_{\mu\nu})_{0\leq \mu,\nu\leq
  3}$, which was proven in \cite{eta_Paper} to be similar to the
partial transpose matrix of $\rho$ up to a factor 1/2.  
It is well-known that the partial transpose criterion is a necessary and sufficient separability condition for two qubits \cite{Peres,Horo}, hence positivity of $M_1(y)$ suffices to prove separability.

In Theorem 4.7 of \cite{FialkowNie2009} the authors solved the $K$-tms
problem of degree 2 in the case where $K$ is defined by a single
quadratic equality, by direct proof rather than using the above
theorems on generic tms. The key point is a result from
\cite{SturmZhang2003}. Applying this theorem to a tms $y$ of degree 2 when $K$ is a sphere, the necessary and sufficient conditions for $y$ to admit a representing measure are $M_1(y)\leq 0$ and $y_{000}-y_{200}-y_{020}-y_{002}=0$. 
Using the mapping between the tms problem and
the separability problem, this theorem of \cite{FialkowNie2009}
directly yields the necessary and sufficient condition $M_1(y)\geq 0$ mentioned above for separability of a symmetric two-qubit state (the condition $y_{000}-y_{200}-y_{020}-y_{002}=0$ being fulfilled for any symmetric two-qubit state). Actually, this problem also coincides
with problem of characterizing the convex hull of spin coherent
states. For spin-1, a necessary and sufficient
criterion was 
established in terms of positivity of a matrix
\cite{GirBraBra08}. Again, this criterion can be shown to coincide with the condition $M_1(y)\geq 0$. Moreover, it was shown in
\cite{KusBen09} that any separable symmetric two-qubit state could be
decomposed as 
a mixture of four pure product states. The tms approach provides a concise
constructive proof of the same fact, as we show in Appendix \ref{app2}. 
 
\begin{table}[t!]
\centering
\begin{tabular}{cccccccccccc}
States \textbackslash $N$          & \textbf{2}               & \textbf{3}             & \textbf{4}               & \textbf{5}             & \textbf{6}               & \textbf{7}             & \textbf{8}               & \textbf{9}             & \textbf{10}               & \textbf{11}             & \textbf{12}               \\ \cline{2-12} 
\multicolumn{1}{c|}{$\rho_{ent}$} & \multicolumn{1}{c|}{0.2} & \multicolumn{1}{c|}{0.2} & \multicolumn{1}{c|}{0.4} & \multicolumn{1}{c|}{0.6} & \multicolumn{1}{c|}{1.0} & \multicolumn{1}{c|}{2.1} & \multicolumn{1}{c|}{5.2} & \multicolumn{1}{c|}{11.6} & \multicolumn{1}{c|}{26.8}  & \multicolumn{1}{c|}{54.6}   & \multicolumn{1}{c|}{170.5}  \\ \cline{2-12} 
\multicolumn{1}{c|}{$\rho_{sep}$}  & \multicolumn{1}{c|}{0.7} & \multicolumn{1}{c|}{0.4} & \multicolumn{1}{c|}{0.6} & \multicolumn{1}{c|}{1.0} & \multicolumn{1}{c|}{2.0} & \multicolumn{1}{c|}{4.2}  & \multicolumn{1}{c|}{10.2} & \multicolumn{1}{c|}{20.8} & \multicolumn{1}{c|}{66.9} & \multicolumn{1}{c|}{94.5} & \multicolumn{1}{c|}{716.3} \\ \cline{2-12} 
\end{tabular}
\caption{Timing of the algorithm in seconds for $N$-qubit symmetric states as function of $N$, averaged over 100 random states, when run on a standard
  desktop PC. 
 The first row corresponds
  to random states drawn from the uniform Haar measure (following
  \cite{Zyczkowski}), which are usually entangled. They are typically
  detected by the condition $M_k(y)\geqslant 0$. The second row
  corresponds to random separable states created by randomly mixing
  random pure separable states. The timing can vary depending on the
  separable state tested and the randomly generated functional $R$ in
  \eqref{SDPGeneral}. 
Up to six different $R$ are tested before
  moving to the next order.}  
\label{table:timing}
\end{table}

\subsection{$N$-qubit symmetric states}
\label{sec:tmssymmetric}

The case of an $N$-qubit symmetric state $\rho$ can be mapped onto the tms problem of Eq.~\eqref{defIntergalSymCompact} where $\vec{x}=(x_1,x_2,x_3)$ is a vector of $\mathbb{R}^3$. We define $K=\{x \in \mathbb{R}^3| g(\vec{x})=0\}$, with $g(\vec{x})=x_1^2+x_2^2+x_3^2-1$ as in the two-qubit case. The highest degree of the monomial $x^{\alpha}$ in \eqref{defIntergalSymCompact} is the total number of indices of the tensor $X_{\mu_1\ldots \mu_N}$, i.e.~$d=N$. The degree of the polynomial defining $K$ is 2, and therefore $d_0=1$. 

To numerically investigate the algorithm for an $N$-qubit symmetric
state we have to solve a SDP with 
degree $d=N$ and flatness condition rank$M_k(z)=$rank$M_{k-1}(z)$. 
 If the state is entangled ($\rho_{ent}$) the SDP
 \eqref{SDPGeneral}--\eqref{SDPGeneral3} should prove infeasible at
 some value of $k$, but this usually happens already at the lowest
 order $k=k_0$. {When the state is separable ($\rho_{sep}$) the
   algorithm has to find a flat extension for some 
$k$, which may require to run the SDP for more than one $R$, or to
increase the values of $k$. Hence, the run time is typically longer
than in the case of an entangled state, as can be seen in 
Table \ref{table:timing}. Usually 
we found a flat extension 
either  at the lowest order $k=k_0$ or 
at order $k=k_0+1$. } 

\subsection{Physical interpretation of the positivity of $M_k(y)$}\label{phys}
Consider a $2k$-qubit symmetric state $\rho$. 
The necessary   condition $M_k(y)\geq 0$ of
Sec.\ref{subsec:maths} turns out to be  equivalent to the positivity of the 
partial transpose of $\rho$ with respect to the $k$ first qubits.  
Indeed, let $T$ be the real symmetric matrix
defined by 
\begin{equation}
\label{etaGeneral}
T_{\bfmu,\bfnu}= X_{\mu_1\dots\mu_{k}\nu_1\ldots\nu_k}
\end{equation}
in terms of the coordinates $X_{\mu_1\mu_2...\mu_{2k}}$ of $\rho$
[see Eq.~\eqref{projrho}], where matrix indices $\bfmu$ and $\bfnu$ are multi-indices
$\bfmu=(\mu_{1},\ldots,\mu_k)$ and $\bfnu=(\nu_{1},\ldots,\nu_{k})$,
with $0\leq\mu_i,\nu_i\leq 3$. Then,
up to a constant numerical factor, the matrix $T$ is similar to the partial
transpose of the density matrix in the computational basis
for the partition
into two sets of $k$ qubits each \cite{eta_Paper}.  
Moreover, $T$ has some recurring rows
and columns, which when removed yield exactly
the moment matrix $M_k(y)$. A symmetric matrix is positive
semi-definite if and only if all principal minors, i.e.~the
determinant of all submatrices, are non-negative.  The determinant of a
matrix which has a recurring column or row is equal to zero, so only
the submatrices with non-recurring rows and columns have to be
considered. Therefore, $T$, and thus the partial transpose, is positive
semi-definite if and only if the matrix $M_k(y)$ is positive
semi-definite.  
So the necessary condition $M_k(y)\geq 0$ is equivalent to the
positive partial transpose criterion of a symmetric state of
$2k$-qubits with equal size partitions. Since for a separable
$N$-qubit state $\rho$ 
any reduced density matrix of $2k$ qubits has to be separable, the
necessary conditions $M_k(y)\geq 0$ with $k\leq N/2$
can be interpreted as positivity of the partial transpose of the
reduced density matrices of $\rho$. This
provides an interesting interpretation of the 
physical meaning of the positivity of the moment matrix.

\subsection{Minimal number of pure
product states needed}
\label{sec:}
If a quantum state is separable it can be written as a convex sum of
product states. Replacing each product state by its eigenvalue-eigenvector decomposition we obtain a decomposition of the initial quantum state as a convex sum of pure product states. What is the minimal number $r_{\textrm{min}}$ of pure product states required to decompose an arbitrary separable state? 

The answer is unknown in the general case. 
For symmetric states, pure states in the decomposition have to be
symmetric themselves (see e.g.~Theorem 1 in \cite{eta_Paper}). As the
appendix \ref{app2}
shows, and as was obtained in  \cite{KusBen09}, in the case of two
qubits, four states are sufficient to represent any separable
symmetric state. 

The above algorithm yields rank$M_k(z)=r$ as an upper bound to the number of pure states required to decompose a given quantum state. In order to investigate systematically the number of states required, we generated symmetric separable states by mixing a large number $m$ of random separable symmetric pure states with random weights as
\begin{equation}
\label{rhosepsym}
\rho_{\textrm{sep,sym}}=\sum_{i=1}^m w_i \left(\ket{\psi_i}\bra{\psi_i}\right)^{\otimes N},
\end{equation} 
with $\sum_i w_i=1$, and applied the algorithm 
to the resulting mixed states. When our
algorithm stops with a flat extension $z$ such that rank$M_k(z)=r$,
then $r$ is an upper bound on the true minimal number of separable
states required to express $\rho_{\textrm{sep,sym}}$. Indeed, since
the extension depends on the random choice of $R_\alpha$ there may be
extensions with a smaller rank, as the algorithm does not minimize
this rank. Therefore every number $r<m$ obtained should give an
           upper bound to the actual generic value for
           $r_{\textrm{min}}$. 
In practice we generated a large list of separable symmetric states
with a value of $m =25$ for $N \leqslant 6 $ and $m=45$ for $N>6$ and
found a flat extension for each one. The smallest numbers found are
reported in Table \ref{table:mindeco}.  

\begin{table}[t!]
\centering
\begin{tabular}{|c|c|c|c|} 
\hline
\textbf{$N$} & \textbf{Min $r$} & \textbf{\# min}	 & \textbf{States tested}  \\ \hline
2            & 4              	&  37304  & 61494                   \\ \hline
3          & 6                &  2410   & 60641                   \\ \hline
4            & 9         	    &  1104   & 174011                  \\ \hline
5          & 12           	&   17     & 174193                  \\ \hline
6            & 17               &	408    & 153081                  \\ \hline
7          & 22            	&   18     & 16129                   \\ \hline
8            & 29              	&	 12     & 16030                    \\ \hline
9            & 35              	&	 2     & 10000                    \\ \hline
10            & 42              	&	 1     & 10000                    \\ \hline
\end{tabular}
\caption{The smallest value of $r$ found, which gives an upper bound
  on the true value $r_{\textrm{min}}$ of the maximal number of pure
  states needed to generate every separable symmetric state. In the
  third column, \# min gives the
   number of states for which the value min $r$ has been reached among
   the states tested. 
 }
\label{table:mindeco}
\end{table}

\section{A new solution to a particular tms problem}
\label{sec:newsol}

The mapping presented above not only helps solving the separability
problem, but it can also, conversely, shed light on particular tms
problems by using results from entanglement theory. We now
give an example of such a situation. 

One of the best-known results from  entanglement theory is the
Peres-Horodecki criterion, which  states that $2\times 2$ and $2\times
3$ systems are separable if and only if the partial transpose is
positive \cite{Peres,Horo} (PPT-criterion). It has been generalized to
the following two statements:
If $\rho$ is supported on $\bbc^2\times\bbc^N$ and the rank 
$r(\rho)= N$ then
$\rho$ is separable (Theorem 1 of \cite{KraCirKarLew00}) and can be
written as a convex sum of projectors on $N$ product 
vectors (Corollary 3a of \cite{KraCirKarLew00}). \\

When $\rho$ is fully symmetric, the above characterizations yield the
following result: 
\begin{theorem}[\cite{EckSchBruLew02}]
\label{th2}
Let $\rho$ be a symmetric $N$-qubit state with positive partial
transpose with respect to the first qubit. If $N=2$ or $3$, or if
$N>3$ and $r(\rho)\leq N$,  then $\rho$ is fully separable.
\end{theorem}
Note that for four qubits there exist entangled symmetric states with a positive partial transpose \cite{TurAugHylKusSamLew12}.
As shown in  \cite{eta_Paper}, the PPT conditions can be
expressed as linear matrix inequalities involving the entries of the
tensor $X_{\mu_1\mu_2\ldots\mu_N}$, or equivalently the
$y_\alpha$. Rewriting the above theorem for $N=3$ 
in the language of $K$-tms
problems, this yields a theorem for a special case of a tms. Even
more, by using the fact that $2\times 3$ systems are separable if and
only if they are PPT, 
we directly get a necessary and sufficient condition for a tms problem
of degree $d=3$ to admit a representing measure supported on the unit
sphere of $\mathbb{R}^3$. This condition reads 
\begin{widetext}
\begin{normalsize}
\begin{equation}
\label{CNS32}
\left(
\begin{array}{cccccccc}
 y_{000}+y_{001} & y_{100}-\mathrm{i} y_{010} & y_{100}+y_{101} &
   y_{200}-\mathrm{i} y_{110} & y_{010}+y_{011} & y_{110}-\mathrm{i} y_{020} &
   y_{001}+y_{002} & y_{101}-\mathrm{i} y_{011} \\
 y_{100}+\mathrm{i} y_{010} & y_{000}-y_{001} & y_{200}+\mathrm{i} y_{110} &
   y_{100}-y_{101} & y_{110}+\mathrm{i} y_{020} & y_{010}-y_{011} &
   y_{101}+\mathrm{i} y_{011} & y_{001}-y_{002} \\
 y_{100}+y_{101} & y_{200}-\mathrm{i} y_{110} & y_{200}+y_{201} &
   y_{300}-\mathrm{i} y_{210} & y_{110}+y_{111} & y_{210}-\mathrm{i} y_{120} &
   y_{101}+y_{102} & y_{201}-\mathrm{i} y_{111} \\
 y_{200}+\mathrm{i} y_{110} & y_{100}-y_{101} & y_{300}+\mathrm{i} y_{210} &
   y_{200}-y_{201} & y_{210}+\mathrm{i} y_{120} & y_{110}-y_{111} &
   y_{201}+\mathrm{i} y_{111} & y_{101}-y_{102} \\
 y_{010}+y_{011} & y_{110}-\mathrm{i} y_{020} & y_{110}+y_{111} &
   y_{210}-\mathrm{i} y_{120} & y_{020}+y_{021} & y_{120}-\mathrm{i} y_{030} &
   y_{011}+y_{012} & y_{111}-\mathrm{i} y_{021} \\
 y_{110}+\mathrm{i} y_{020} & y_{010}-y_{011} & y_{210}+\mathrm{i} y_{120} &
   y_{110}-y_{111} & y_{120}+\mathrm{i} y_{030} & y_{020}-y_{021} &
   y_{111}+\mathrm{i} y_{021} & y_{011}-y_{012} \\
 y_{001}+y_{002} & y_{101}-\mathrm{i} y_{011} & y_{101}+y_{102} &
   y_{201}-\mathrm{i} y_{111} & y_{011}+y_{012} & y_{111}-\mathrm{i} y_{021} &
   y_{002}+y_{003} & y_{102}-\mathrm{i} y_{012} \\
 y_{101}+\mathrm{i} y_{011} & y_{001}-y_{002} & y_{201}+\mathrm{i} y_{111} &
   y_{101}-y_{102} & y_{111}+\mathrm{i} y_{021} & y_{011}-y_{012} &
   y_{102}+\mathrm{i} y_{012} & y_{002}-y_{003} \\
\end{array}
\right) \geqslant 0.
\end{equation}
\end{normalsize}
\end{widetext}
(see the expression of  \cite{eta_Paper} which explicitly gives the
PPT criterion for 3 qubits). 
This result does not appear to have been known previously in the tms
literature.\\

While \eqref{CNS32} is a necessary and sufficient condition in the case of a tms of degree $N=3$, Theorem \ref{th2} also provides us with a sufficient condition for a tms of arbitrary degree $N$. Indeed, suppose one wants to know whether a given tms $y$ of degree $N>3$ admits a representing measure on the unit sphere. Using the mapping inverse to the one in Sec.\ref{sec:tmssymmetric} one can construct the density matrix $\rho$ associated with the tms via \eqref{projrho}.  If $\rho$ is PPT and has rank $r(\rho)\leq N$, then  there exists a representing measure.

\section{Conclusions}
We have proposed a new and elegant solution of the entanglement
problem by mapping it to the truncated $K$-moment problem. Benefiting
from the mathematically well-developed field of the 
theory of moments, we provide an algorithm that for an entangled state
certifies its entanglement in a finite
number of steps. If the state is separable,
it usually halts at the first iteration ($k=k_0$ in
Fig.\ref{fig:Flow}) and then returns an
explicit decomposition of the state into a convex  
sum of product states.  
Similarly to previous algorithms, our algorithm makes use of
semi-definite programming and
``extensions'', but there are a number of conceptual differences that
allow us express and solve the problem very elegantly and adapt it
easily to different physical situations, including subsystems of
different dimensions or symmetries, or incomplete data. \\

In our approach, rather than 
working directly with the density matrix,
the semi-definite optimization problem is based on moment matrices and
localizing matrices, where the latter incorporate the
constraints of the states of the sub-spaces.  This is possible since
these states in the sub-spaces are 
restricted to compact sets characterized by polynomial constraints (e.g.~to Bloch spheres in the case of individual spins-1/2).  Both the moment
matrix and the localizing
matrices must be positive semi-definite for a state to be
separable. 
Extensions are extensions 
of the moment matrix, and we need not impose a particular symmetry
on such an extension, nor positivity of the partial transpose of the
state, since this is taken care of by positivity of the moment matrix
(see Sec.\ref{phys}).

Our algorithm contains in addition a crucial element, namely
the idea of ``flat extensions'': if at a given order $k$ of the extension
the SDP is feasible, one checks whether the rank of the extended
moment matrix is the same as the one at order $k-d_0$ [with $d_0$
related to the largest degree of the constraint polynomials, see after
eq.\eqref{flatcond}]. If so, the state is separable and one obtains its
explicit convex decomposition into product states. 
In  \cite{NaOwPl09} it was already noted
that when PPT is imposed on the extensions in the algorithm by Doherty
et al.~\cite{DohParSpe02,DoPaSp04,DoPaSp05}, {\em sometimes} 
separability can be concluded in a finite number of steps by checking
whether the rank of the found extension of the density matrix has not
increased compared to 
the original state, a situation called ``rank loop''.  There, the sufficiency
of a rank loop for separability follows from a theorem due to
Horodecki et al.~\cite{HoLeViCi00}, 
according to which a PPT state is
separable if its rank is smaller or equal than the rank of the reduced
state. In our case, the implementation of the flat-extension query is
a decisive part of the algorithm, based on Theorem \ref{th1}. \\

 Formulating the entanglement as a truncated $K$-tms problem also has the
 advantage that the algorithm readily accepts incomplete data from an
 experiment. Indeed, since for multi-partite systems fully determining
 the state requires an effort that grows exponentially with the number
 of subsystems, fully specifying or experimentally determining the
 state becomes at some point
 impossible in practice. Since our algorithm is based from the very
 beginning on a truncated sequence of 
 moments (that can be chosen to be expectation values of Hermitian
 operators that {\em were} measured), we can leave open additional
 moments that were 
 not measured and still run the algorithm. Using the algorithm in this
 way should allow one 
 to determine how many and which 
 moments one should measure in order to still be able to prove that a
 state is entangled. \\

Finally, as symmetric states of $N$ qubits coincide with spin-$j$
states with $N=2j$, separable symmetric states can be identified with
classical spin-$j$ states (see
e.g.~\cite{GirBraBra08}). The latter, 
defined in \cite{GirBraBra08}, are convex combinations of
spin-coherent states, and can be considered the quantum states which
are closest to  having a classical behaviour in the sense of minimal quantum
fluctuations \cite{GirBraBra12,PRL_Tensor,eta_Paper}.  Applying the
algorithm presented here to symmetric states of $N$-qubits also allows
one to check whether a spin-$j$ state is classical.

\section*{Acknowledgments}
 D.B.~thanks O.G., the LPTMS, LPS, and the
Universit\'e Paris Saclay for 
hospitality. We thank the Deutsch-Franz\"osische 
Hochschule (Universit\'e franco-allemande) for support, grant
number CT-45-14-II/2015. Ce travail a b\'en\'efici\'e
d'une aide Investissements d'Avenir du LabEx PALM
(ANR-10-LABX-0039-PALM).

\appendix

\section{Matlab implementation}
\label{app1}
Here we give a Matlab implementation of the easiest case of the
symmetric state of two qubits, (or a spin-1 state
\cite{quantumnessspin1}). The quantum state $\rho$ is given as in
\eqref{Tensorrep} as 
\begin{equation}
X_{\mu_1\mu_2}=\tr\{ \rho \, P^{\dagger}_s \sigma_{\mu_1} \otimes
\sigma_{\mu_2} P_s \}, \label{eq.40}
\end{equation}
with $P_s$ the projector onto the symmetric states. The following
implementation uses Matlab and the programs GloptiPoly 3
\cite{GloptiPoly3} and the solver SeDuMi \cite{SeDuMi}. {To
  increase the 
probability of finding a flat extension, the semi-definite solver
should use the highest possible accuracy in the calculation of the
minimal value of the SDP. }

\begin{lstlisting}
mpol x 3
xMom=[1 x(1) x(2) x(3) x(1)^2  x(1)*x(2) x(1)*x(3) x(2)^2 x(2)*x(3) x(3)^2];
y=[X(1,1) X(1,2) X(1,3) X(1,4) X(2,2) X(2,3) X(2,4) X(3,3) X(3,4) X(4,4)];
con=[mom(xMom)==y];
K = [x(1)^2+x(2)^2+x(3)^2-1==0];
G = randn(length(xMom));
R = xMom*(G'*G)*xMom'; k=2;
P = msdp(min(mom(R)),K,con,k);
pars.eps=0; mset(pars);
[status] =msol(P);
\end{lstlisting}
Line 3 is given by Eq.~\eqref{eq.40}.
Line 4 corresponds to Eq.\eqref{defIntergalSymCompact}. K fixes the
variables to Bloch vectors of length 1. R is the arbitrary positive
bounded polynomial which should be minimized. At line 8, 'msdp'
formulates the problem in the language of SDPs
(construction of moment matrices and localizing matrices).  Line 9
sets the accuracy of the SDP solver to its highest value. At line 10,
'msol' solves the SDP. 

\begin{itemize}

\item If the problem is detected as infeasible (status=-1) the state
  is entangled.  

\item If there is no flat extension found (status=0), one can re-run
  the program with a different R, or increase the order by one.   

\item If the state is separable and a flat extension is found
  (status=1) the solution can be extracted with the command
  "sol=double(x)". Then "sol" contains a list of Bloch vectors of the
  pure states that give a decomposition into separable states as in
  Eq.~\eqref{tensorseparablestatesDef}. The vector of weights $w_i$
  can then be easily calculated.

\end{itemize}

This implementation can be extended to a larger number of
qubits by adapting the monomial basis in line 2 to a higher degree and
line 3 to contain all entries of the tensor $X_{\mu_1\ldots\mu_N}$
(Eq.~\eqref{Tensorrep}).   
The 
generalization to non-symmetric states is also possible, but the
number of variables 
increases. E.g.~two qubits would require one independent Bloch vector
for each subsystem, so one would need six variables in total.

\section{Minimal rank for symmetric two-qubit states}
\label{app2}

Theorem 4.7 of \cite{FialkowNie2009} states that a tms $y$ of degree 2
admits a representing measure supported by $K$ if and only if
$M_1(y)$ is positive and
$y_{000}-y_{200}-y_{020}-y_{002}=0$. We therefore obtain that
a two-bit symmetric state $\rho$ is separable if and only if it is
associated with a tms such that $M_1(y)$ is positive and
$y_{000}-y_{200}-y_{020}-y_{002}=0$. These two conditions in
fact coincide respectively with the PPT criterion (see Sec.\ref{phys}) and with the
condition that $X_{00}=\sum_{a=1}^3X_{aa}$. The
latter condition 
itself is a consequence of properties of the projections of tensor
products of Pauli matrices over the symmetric subspace, as was shown
in \cite{PRL_Tensor}.

The proof of the fact that iff $M_1(y)$ is positive and
$y_{000}-y_{200}-y_{020}-y_{002}=0$ then $\rho$ is separable
into a mixture of only 4 separable states can be simplified by 
using the tms formalism. Let us derive the necessary and sufficient
condition above in our language. The 'necessary' direction is
obvious. The proof for the 'sufficient' direction goes as follows. Let
us assume that the coordinates $X_{\mu\nu}$ form a positive rank-$r$
matrix $M_1(y)$.
Since the state is symmetric, $M_1(y)$ is a real symmetric
4$\times$ 4 matrix and hence $r\le 4$.  
Then $M_1(y)$ can be decomposed into a sum of $r$ 
projectors 
on orthogonal 
vectors $u^{(k)}$ as
\begin{equation}
\label{4proj}
X_{\mu\nu}=\sum_{k=1}^r u^{(k)}_\mu u^{(k)}_\nu.
\end{equation}
Let $\Delta_{u}=(u_0)^2-\sum_{a=1}^3(u_a)^2$ for any 4-vector
$u$. Since $X_{\mu\nu}$ verify $X_{00}=\sum_{a=1}^3X_{aa}$ we have
$\sum_{i=1}^r \Delta_{u^{(i)}}=0$. 
Whenever $\Delta_{u^{(i)}}=0$, one has $u^{(i)}_0\neq 0$ (otherwise
the whole vector $u^{(i)}$ vanishes and does not contribute to the sum
\eqref{4proj}), 
so that the corresponding projector can be rewritten 
\begin{equation}
\label{uunn}
u^{(i)}_\mu u^{(i)}_\nu=\left(u^{(i)}_0\right)^2 n^{(i)}_\mu n^{(i)}_\nu
\end{equation}
with $n^{(i)}=(1,{\bf n})$ and $|{\bf n}|=1$. If all
$\Delta_{u^{(i)}}=0$ then Eqs.~\eqref{4proj}--\eqref{uunn} immediately
yield a sum over $r$ separable pure states. If not, then since $\sum_i
\Delta_{u^{(i)}}=0$ there must be two indices $i$ and $j$ with
$\Delta_{u^{(i)}}<0$ and $\Delta_{u^{(j)}}>0$. Let $v(t)=t u^{(i)}
+(1-t)u^{(j)}$. Then $\Delta_{v(0)}>0$ and $\Delta_{v(1)}<0$, so that
there has to be a $t_c\in ]0,1[$ such that $\Delta_{v(t_c)}=0$. The
vector $v'(t)=-(1-t) u^{(i)} + t u^{(j)}$ is then such that  
\begin{equation}
u^{(i)}_\mu u^{(i)}_\nu+u^{(j)}_\mu u^{(j)}_\nu=
\frac{v(t_c)_\mu v(t_c)_\nu+v'(t_c)_\mu
  v'(t_c)_\nu}{t_c^2+(1-t_c)^2}. 
\end{equation}
Then subtracting a projector on $v(t_c)$ yields
\begin{equation}
X_{\mu\nu}-\frac{v(t_c)_\mu v(t_c)_\nu}{t_c^2+(1-t_c)^2}  
=\sum_{k=1}^{r-1} \tilde{u}^{(k)}_\mu \tilde{u}^{(k)}_\nu
\end{equation}
where $\tilde{u}^{(k)}$ are the orthogonal states $u^{(k')}$ ($k'\neq
i,j$) and $v'(t_c)$. Because of the definition of $t_c$ and using
\eqref{uunn}, the projector on $v(t_c)$ is proportional to a projector
representing a separable pure state, and the remaining sum is such that
$\sum_{k=1}^{r-1} \Delta_{\tilde{u}^{(k)}}=0$. We are therefore back
to the form \eqref{4proj} but with the rank reduced by one. The
same procedure can be applied repeatedly to further reduce the rank
down to 1; the last projector is then necessarily of the form
\eqref{uunn}. In the end, $\rho$ is written as a sum of $r\leq 4$
projectors on separable pure states.




\end{document}